\documentstyle[prb,aps,twocolumn,epsf]{revtex}

\begin{document}
\draft
\title{Renormalization group approach to Fermi Liquid Theory }
\author{
N. Dupuis$^{(1,2)}$ and G. Y. Chitov$^{(1)}$ }
\address{
$^{(1)}$D\'epartement de Physique et Centre de Recherche en Physique du
Solide, \\
Universit\'e de Sherbrooke, Sherbrooke, Qu\'ebec, Canada J1K 2R1. \\
$^{(2)}$Laboratoire de Physique des Solides,
Universit\' e Paris-Sud,
91405 Orsay, France   }
\date{November 24, 1995}
\maketitle
\begin{abstract}
We show that the renormalization group (RG) approach to interacting fermions
at one-loop order recovers Fermi liquid theory results when the forward
scattering zero sound (ZS) and exchange (ZS$'$) channels are both taken into
account. The Landau parameters are related to the fixed point value
of the ``unphysical'' limit of the forward scattering vertex. We specify
the conditions under which the results obtained at one-loop order hold at
all order in a loop expansion. We also emphasize the similarities between
our RG approach and the diagrammatic derivation of Fermi liquid theory.  
\end{abstract}

\narrowtext
\vspace{.4cm} 

Much of our understanding of interacting fermions is based on 
the Fermi liquid theory (FLT). \cite{Landau57,Landau59} Although the latter 
was first formulated as a phenomenological theory, its microscopic
foundation was rapidly established using field theoretical
methods. \cite{Landau59,Abrikosov63}  The discovery of new
materials showing non Fermi liquid behavior, like high-$T_c$
superconductors, has motivated a lot of theoretical work in
order to clarify the validity and the limitations of the FLT.

Several authors have recently applied renormalization group (RG)
methods to interacting fermions (see 
[\onlinecite{Benfatto90,Shankar91,Polchinski92,Chitov95}] 
and references therein). 
While these methods are well known in the context of
one and quasi-one dimensional interacting fermions systems where they have
been very successful,\cite{Solyom79} their
application to isotropic systems of dimension $d$
greater than one is more recent. In his study of
interacting fermions in $d=2,3$, Skankar used both RG methods and standard
perturbative calculation.\cite{Shankar91} While RG arguments were used to
identify the relevant couplings, the low-energy degrees of freedom were
explicitly integrated out in the Landau interaction channel by means of 
standard diagrammatic calculations. Extending Shankar's approach to finite
temperature, Chitov and S\'en\'echal 
have recently shown how this interaction channel can be treated by RG
without use of any additional perturbative
calculation.\cite{Chitov95} Moreover, the finite temperature formalism
clearly establishes the difference between the ``physical'' and
``unphysical'' limits of the forward scattering vertex and therefore 
differentiates the Landau function 
(i.e. the Landau parameters $F^s_l$ and $F^a_l$) from the (physical) forward
scattering amplitude. It is clear that both approaches
\cite{Shankar91,Chitov95} amount to suming the series of bubble
diagrams in the forward scattering zero sound  channel. Since it is well
known that such an RPA-type calculation reproduces the results of FLT,
\cite{Mahan90} the agreement between FLT and RG approach is a posteriori
not surprising. Although the selection of Feynman diagrams appearing in the
RG procedure was justified on
the basis of an expansion in the small parameter $\Lambda _0/K_F$ ($K_F$ is
the Fermi wave-vector and $\Lambda _0$ a low-energy cut-off), it is
nevertheless rather 
unexpected that the RG approach reduces to an RPA calculation while
the diagrammatic microscopic derivation of FLT \cite{Landau59,Abrikosov63} is
obviously more than a simple RPA calculation.

The aim of this paper is to reconsider the RG approach to interacting
fermions along the lines developed in Refs.\
[\onlinecite{Shankar91,Chitov95}] in 
order to clarify its connection with FLT. First, we derive the RG
equation for the ``physical'' (``unphysical'')  limit of the forward
scattering  vertex $\Gamma ^Q$ ($\Gamma ^\Omega $) at one-loop
order. In order to respect the Fermi statistics, the forward scattering
zero sound (ZS) and exchange (ZS$'$) channels are
both taken into account. As a result, we find that both the flows of $\Gamma
^Q$ and $\Gamma ^\Omega $ are non zero. We show that the antisymmetry of
$\Gamma ^Q$ under exchange of the two incoming or outgoing particles is
conserved under RG, while the antisymmetry of $\Gamma ^\Omega $ is lost. We
then solve (approximately) the RG equations to obtain a relation between
the fixed point (FP) values ${\Gamma ^Q}^*$ and ${\Gamma ^\Omega }^*$. The
standard relation between ${\Gamma ^Q}^*$ and the Landau parameters $F^s_l$,
$F^a_l$ (which is one of the key results of the microscopic diagrammatic
derivation of FLT) is recovered if one identifies these latter with ${\Gamma
^\Omega }^*$. This result differs from previous RG approaches
\cite{Shankar91,Chitov95} where the Landau parameters were identified with
the bare interaction of the low-energy effective action (which is the
starting point of the RG analysis).  
We show that the relation between ${\Gamma ^\Omega }^*$ and
the Landau parameters obtained at one-loop order holds at all order if one
assumes that the only singular contribution to the RG flow is due to the
one-loop ZS graph.

We consider a two-dimensional system of interacting spin one-half fermions
with a circular Fermi surface (the results obtained in this paper can be
straightforwardly extended to the three-dimensional case). 
Following Refs.\ [\onlinecite{Shankar91,Chitov95}], we write the partition 
function
as a functional integral over Grassmann variables, $Z=\int {\cal D}\psi^*
{\cal D}\psi e^{-S}$, where $S$ is a low-energy effective action 
(we set $\hbar =k_B=1$): 
\begin{eqnarray}
S &=& -\sum _{\tilde K,\sigma } \psi ^*_\sigma (\tilde K)(i\omega -\epsilon
({\bf K})+\mu  ) \psi _\sigma (\tilde K) \nonumber \\
& & +{1 \over {4\beta \nu}} \sum _{\tilde K_1...\tilde K_4} \sum _{\sigma
_1... \sigma _4} U_{\sigma _1\sigma _2,\sigma _3\sigma _4} ({\bf K}_1,{\bf
K}_2,{\bf K}_3, {\bf K}_4)
\nonumber \\ & & \times  
\psi ^*_{\sigma _4}(\tilde K_4)
\psi ^*_{\sigma _3}(\tilde K_3)
\psi _{\sigma _2}(\tilde K_2)
\psi _{\sigma _1}(\tilde K_1)
\nonumber \\ & & \times 
\delta _{{\bf K}_1+{\bf K}_2,{\bf K}_3+{\bf K}_4}
\delta _{\omega _1+\omega _2,\omega _3+\omega _4}+\cdot \cdot \cdot \,,
\label{action}
\end{eqnarray}
where the dots denote terms which are irrelevant at tree-level.
\cite{Shankar91} Here ${\bf K}$ is a two-dimensional vector with 
$\vert K-K_F \vert <\Lambda _0\ll K_F$. $\mu $ is the chemical potential, $K_F$
the Fermi wave-vector and the cut-off $\Lambda _0$ fixes the energy scale of
the effective action. $\tilde K=(\bf K,\omega )$ and $\omega $ is a
fermionic Matsubara frequency. $\beta =1/T$ is the inverse temperature and
$\nu $ the size of the system. $\sigma =\uparrow ,\downarrow $ refers
to the electron spins. The antisymmetrized coupling function $U_{\sigma
_1\sigma _2,\sigma _3\sigma _4} ({\bf K}_1,{\bf K}_2,{\bf K}_3, {\bf K}_4)$
is assumed to be a non-singular function of its arguments. 
Ignoring irrelevant terms, we write the single particle energy as 
$\epsilon ({\bf K})=\mu +v_Fk$ where $K=K_F+k$ and $v_F$ is the Fermi velocity.
The summation over the wave vectors is defined by
\begin{equation}
{1 \over \nu } \sum _{\bf K}=\int {{d^2{\bf K}} \over {(2\pi )^2}} 
\equiv K_F \int _{-\Lambda _0}^{\Lambda _0} {{dk} \over {2\pi }} 
\int _0^{2\pi } {{d\theta } \over {2\pi }} \,,
\end{equation}
keeping only the relevant term in the integration measure at tree-level. 
Shankar's analysis of the coupling
functions of the quartic interaction shows that only two such functions
survive under the RG flow for $\Lambda _0\ll K_F$: the forward scattering
coupling function and the BCS coupling function. In the following, we
neglect this latter by assuming it is irrelevant at one-loop order so that 
no BCS instability occurs. The forward scattering coupling function is
denoted by $\Gamma _{\sigma _i}(\tilde K_1,\tilde K_2,\tilde K_2-\tilde
Q,\tilde K_1+\tilde Q)$ where $\tilde Q=({\bf Q},\Omega )$ with $Q\ll K_F$
and $\Omega $ is a bosonic Matsubara frequency 
(we use the notation $\Gamma _{\sigma
_i}\equiv \Gamma _{\sigma _1\sigma _2,\sigma _3\sigma _4}$). Since the
dependence on $k_{1,2}$ and $\omega _{1,2}$ is irrelevant, we introduce the
coupling function $\Gamma _{\sigma _i}(\theta _1,\theta _2,\tilde Q)=
\Gamma _{\sigma _i}({\bf K}^F_1,{\bf K}^F_2,{\bf K}^F_2-\tilde
Q,{\bf K}^F_1+\tilde Q)$ where ${\bf K}^F=K_F{\bf K}/K=K_F(\cos
\theta ,\sin \theta )$ is a wave-vector on the Fermi surface. The forward
scattering coupling function can be decomposed 
into a spin triplet amplitude $\Gamma _t$ and a spin singlet amplitude
$\Gamma _s$: \cite{Baym91,nota}
\begin{eqnarray}
&& \Gamma _{\sigma _i}(\theta _1,\theta _2,\tilde Q) =
{{\Gamma _t  (\theta _1,\theta _2,\tilde Q)} \over 2}
(\delta _{\sigma _1,\sigma _4}\delta _{\sigma _2,\sigma _3}+
\delta _{\sigma _1,\sigma _3}\delta _{\sigma _2,\sigma _4})
\nonumber \\
& & + {{\Gamma _s  (\theta _1,\theta _2,\tilde Q)} \over 2}
(\delta _{\sigma _1,\sigma _4}\delta _{\sigma _2,\sigma _3}-
\delta _{\sigma _1,\sigma _3}\delta _{\sigma _2,\sigma _4}) \,.  
\label{trsu}
\end{eqnarray}
We now introduce the ``physical'' ($\Gamma ^Q$) and ``unphysical'' ($\Gamma
^\Omega $) limits of the forward scattering vertex:
\begin{eqnarray}
\Gamma _{\sigma _i}^Q(\theta _1-\theta _2) &=& \lim _{Q\rightarrow 0} \Bigl 
\lbrack \Gamma _{\sigma _i}(\theta _1,\theta _2,\tilde Q) 
\Bigl \vert _{\Omega =0} \Bigr \rbrack \,, \nonumber \\
\Gamma _{\sigma _i}^\Omega (\theta _1-\theta _2) &=& \lim _{\Omega 
\rightarrow 0} \Bigl  \lbrack \Gamma _{\sigma _i}(\theta _1,\theta _2,\tilde
Q) \Bigl \vert _{Q=0} \Bigr \rbrack \,. 
\end{eqnarray}
$\Gamma _{\sigma _i}^Q$ and $\Gamma _{\sigma _i}^\Omega $ can be decomposed
into singlet and triplet amplitudes according to (\ref{trsu}). The only
remnant of the antisymmetry of $\Gamma _{\sigma _i}$ is then the condition
$\Gamma ^{Q,\Omega }_t(\theta =0)=0$ for the bare vertices. \cite{Chitov95} 

We now derive the RG equation (using the Kadanoff-Wilson approach
\cite{Shankar91}) for the coupling $\Gamma _{\sigma _i}$ when
the cut-off $\Lambda _0$ is reduced according to $\Lambda (t)=
\Lambda _0e^{-t}$. 
Three diagrams have to be considered at one-loop order, corresponding
to the ZS, ZS$'$ and BCS channels. As pointed out in Ref.\
[\onlinecite{Chitov95}], the ZS graph alone does not respect the Fermi
statistics. Indeed, if one exchanges the two incoming or outgoing lines,
the ZS graph transforms into the ZS$'$ graph and vice versa (Fig.\
1). It is therefore necessary to consider the ZS and ZS$'$ graphs on the same
footing. We ignore momentarily the symmetry-preserving contribution of the
BCS diagram which will be discussed later. 

The contribution of the ZS graph is: \cite{Chitov95}
\begin{eqnarray}
&& {{d\Gamma ^Q_{\sigma _i}(\theta _1-\theta _2)} \over {dt}} \Biggl \vert
_{\rm ZS}  = - {{N(0)\beta _R} \over 
{\cosh ^2(\beta _R)}}   \nonumber \\ && 
\times  \int {{d\theta } \over {2\pi }} 
\sum _{\sigma ,\sigma '} 
\Gamma _{\sigma _1\sigma ',\sigma \sigma _4}^Q (\theta _1-\theta )
\Gamma _{\sigma \sigma _2,\sigma _3\sigma '}^Q (\theta -\theta _2) \,, 
\nonumber \\ &&
{{d\Gamma ^\Omega _{\sigma _i}(\theta _1-\theta _2)}  \over {dt}} 
\Biggl \vert _{\rm ZS} = 0 \,, 
\label{ZS1}
\end{eqnarray}
where $\beta _R=v_F\beta \Lambda (t)/2$ is a dimensionless inverse 
temperature and
$N(0)=K_F/2\pi v_F$ the density of states per spin. Since $\lim _{\beta
\rightarrow \infty } (\beta /4)\cosh ^{-2}(\beta x/2)=\delta (x)$, the ZS
graph gives a singular contribution to the RG flow of $\Gamma ^Q$ when
$T\rightarrow 0$. Consider now the contribution of the ZS$'$ graph. 
For a given  $\tilde Q$, this graph involves the quantity (Fig.\ 1)
\begin{eqnarray}
T\sum _\omega G(\tilde K)G(\tilde K+{\bf K}^F_{21}-\tilde
Q) =&&  \nonumber \\
{1 \over 2} 
{{\tanh \Bigl \lbrack {\beta \over 2}\epsilon ({\bf K}+{\bf K}^F_{21}
-{\bf Q}) \Bigr \rbrack  - \tanh \Bigl \lbrack {\beta \over
2}\epsilon ({\bf K}) \Bigr \rbrack }  \over
{-i\Omega +\epsilon ({\bf K})-\epsilon ({\bf K}
+{\bf K}^F_{21}-{\bf Q})}} &&  \,.
\label{ZS'1}
\end{eqnarray}  
Here $G(\tilde K)=(i\omega -v_Fk)^{-1}$ is the one-particle Green's
function and ${\bf K}^F_{21}={\bf K}^F_2-{\bf K}^F_1$. In general, the
limit $\tilde Q\rightarrow 0$ can be taken without
any problem (and is independent of the order in which the limits $Q,\Omega
\rightarrow 0$ are taken) and (\ref{ZS'1}) gives a smooth contribution
to the flow of $\Gamma ^Q$ and $\Gamma ^\Omega $. As pointed out by Mermin,
\cite{Mermin67} problems arise when ${\bf K}^F_{21}$ 
is small since the limits $\tilde Q\rightarrow 0$ and ${\bf K}^F_{21}\to 0$ 
do not commute. For small ${\bf K}^F_2-{\bf K}^F_1$, i.e. for $\vert
\theta _1-\theta _2 \vert \ll T/E_F$, (\ref{ZS'1}) becomes
\begin{equation}
{{v_F{\bf \hat K} \cdot ({\bf K}^F_{21}-{\bf Q})} 
\over {-i\Omega -v_F{\bf \hat K} \cdot ({\bf K}^F_{21}
-{\bf Q})}} {{\beta /4} \over {\cosh ^2(\beta v_Fk/2)}} \,,
\end{equation}
where ${\bf \hat K}={\bf K}/K$ is a unit vector.
This quantity (apart from the thermal factor $\beta \cosh ^{-2}(\beta v_Fk
/2)$) has been analyzed in detail by Mermin who showed that the
antisymmetry of the vertex is strongly related to the order in which the
different limits are taken.   
Following Ref.\ [\onlinecite{Mermin67}], we first take the limit $\tilde
Q\rightarrow 0$ (which is well defined for ${\bf K}^F_1 \neq {\bf
K}^F_2$) and then $\theta _1-\theta _2\rightarrow 0$. This ensures
that $\Gamma ^{Q,\Omega }_{\sigma _i}(\theta )$ is a continuous function at
$\theta =0$. We then have ($\tilde Q=0$):
\begin{equation} 
\lim _{\theta _1 \to \theta _2} \Bigl \lbrack
T\sum _\omega G(\tilde K)G(\tilde K+{\bf K}^F_{21})
 \Bigr \rbrack =
-{\beta \over {4\cosh ^2(\beta v_Fk/2)}} \,.  
\label{ZS'2}
\end{equation}
Eq.\ (\ref{ZS'2}) shows that when $T\rightarrow 0$ the ZS$'$ graph gives a
singular contribution to the RG flow of $\Gamma ^{Q,\Omega }(\theta )$ for
$\vert \theta \vert \ll T/E_F$.\cite{6point}  Taking into account the spin dependence of
the coupling, 
we obtain that the contributions of the ZS and ZS$'$ graphs to the RG flow 
of $\Gamma ^Q_t(\theta =0)$ cancel each over. Consequently,
$\Gamma ^Q_t(\theta =0)=0$  for any value of the flow parameter $t$. The
antisymmetry of $\Gamma ^Q$ is therefore conserved under RG. Since the
contribution  of the ZS graph to the RG flow of $\Gamma ^\Omega (\theta =0)$
vanishes, while the contribution of the ZS$'$ graph does not, the
antisymmetry of $\Gamma ^\Omega $ is not conserved under RG. This
results agrees with standard diagrammatic calculations. \cite{Mermin67} 

Taking into account both the contributions of the ZS and ZS$'$ graphs, the
RG equations of $\Gamma ^{Q,\Omega }$ can be written:
\begin{eqnarray}
{{d\Gamma ^Q_{\sigma _i}} \over {dt}} &=& 
{{d\Gamma ^Q_{\sigma _i}} \over {dt}} \Biggl \vert _{\rm ZS} +
{{d\Gamma ^Q_{\sigma _i}} \over {dt}} \Biggl \vert _{{\rm ZS}'} \,,
\nonumber \\ 
{{d\Gamma ^\Omega _{\sigma _i}} \over {dt}} &=& 
{{d\Gamma ^\Omega _{\sigma _i}} \over {dt}} \Biggl \vert _{{\rm ZS}'} =
{{d\Gamma ^Q_{\sigma _i}} \over {dt}} \Biggl \vert _{{\rm ZS}'} \,. 
\end{eqnarray}
The two preceding equations can be combined (using also (\ref{ZS1})) to 
obtain
\begin{eqnarray}
{{d\Gamma ^Q_{\sigma _i}(\theta _1-\theta _2)} \over {dt}} = 
{{d\Gamma ^\Omega _{\sigma _i}(\theta _1-\theta _2)} \over {dt}}
 - {{N(0)\beta _R} \over 
{\cosh ^2(\beta _R)}}  && \nonumber \\ \times 
\int {{d\theta } \over {2\pi }} 
\sum _{\sigma ,\sigma '} 
\Gamma _{\sigma _1\sigma ',\sigma \sigma _4}^Q (\theta _1-\theta )
\Gamma _{\sigma \sigma _2,\sigma _3\sigma '}^Q (\theta -\theta _2) \,. &&
\label{RGEQ1}
\end{eqnarray}
In order to solve this RG equation, we Fourier transform $\Gamma
^{Q,\Omega }_{\sigma _i}(\theta )$ and introduce the spin symmetric
($A^{Q,\Omega }$) and antisymmetric ($B^{Q,\Omega }$) parts:
\begin{eqnarray}
\Gamma ^{Q,\Omega }_{\sigma _i}(l) &=& \int {{d\theta } \over {2\pi }}
e^{-il\theta } \Gamma ^{Q,\Omega }_{\sigma _i}(\theta ) \,,
\nonumber \\
2N(0)\Gamma ^{Q,\Omega }_{\sigma _i}(l) &=& A^{Q,\Omega }_l 
\delta _{\sigma _1,\sigma _4} \delta _{\sigma _2,\sigma _3} 
+B^{Q,\Omega }_l \mbox {\boldmath $\tau $}_{\sigma _1\sigma _4} \cdot
\mbox {\boldmath $\tau $}_{\sigma _2\sigma _3} \,,
\label{ABdef}
\end{eqnarray}
where $\mbox {\boldmath $\tau $}$ denote the Pauli matrices. 
Eq.\ (\ref{ABdef}) holds 
when the spin dependent part of the particles interaction is due purely to
exchange. Eq.\ (\ref{RGEQ1}) then takes the simple form
\begin{equation}
{{dA^Q_l} \over {dt}} = {{dA^\Omega _l} \over {dt}} 
-{{\beta _R} \over {\cosh ^2(\beta _R)}} {A^Q_l}^2 \,,
\label{RGEQ11}
\end{equation}
and the same equation relating $B^Q_l$ and $B^\Omega _l$.
Integrating these equations between 0 and $t$, we obtain (writing explicitly
the $t$ dependence)
\begin{equation}
A^Q_l(t)=A^\Omega _l(t) - \int _0^t dt' \, 
{{\beta _R} \over {\cosh ^2(\beta _R)}} A^Q_l(t')^2 
\label{RGEQ2}
\end{equation}
and a similar equation for $B^Q_l(t)$. The RG equations in their symmetry
preserving form (\ref{RGEQ11},\ref{RGEQ2}) relate two FP-s ${\Gamma _{\sigma
_i}^Q}^*$ and ${\Gamma _{\sigma _i}^\Omega }^*$ in a fashion more general
than the standard RPA-like form (see Eq.\ (\ref{PV1}) below) with all
harmonics decoupled. Deferring study of such a fixed point, which is beyond
the scope of the present paper, \cite{Chitov} we concentrate now on the
approximation leading to the standard FLT results.
Because of the thermal factor $\beta _R/\cosh ^2(\beta _R)$, the second term
of the rhs of (\ref{RGEQ2}) becomes different from zero only when $\Lambda
(t)\lesssim T/v_F$. On the other hand, we have shown above
that $\Gamma ^\Omega _{\sigma _i}(\theta )$ is a  smooth function of
$\Lambda (t)$ except for $\vert \theta \vert \ll T/E_F$. The Fourier
transform $\Gamma _{\sigma _i}^\Omega (l)$ is also a smooth function of
$\Lambda (t)$. At low temperature, we can therefore make the approximation 
 $A^\Omega _l\vert _{\Lambda (t)\lesssim T/v_F} \simeq {A^\Omega 
_l}^*$, where ${A^\Omega _l}^*=A^\Omega _l\vert _{\Lambda (t)=0}$ is the FP
value of $A^\Omega _l$. This allows us to rewrite (\ref{RGEQ2}) for $\Lambda
(t)\lesssim T/v_F$ as
\begin{equation}
A^Q_l(t)={A^\Omega _l}^*-\int _0^t dt' \, 
{{\beta _R} \over {\cosh ^2(\beta _R)}} A^Q_l(t')^2 \,.
\label{RGEQ3}
\end{equation} 
Eq.\  (\ref{RGEQ3}) is solved by introducing the parameter $\tau =\tanh
\beta _R$. The FP values of $A^Q_l$ and $B^Q _l$ are given in the
zero-temperature limit by
\begin{equation}
{A^Q_l}^*= {{{A^\Omega _l}^*} \over {1+{A^\Omega _l}^*}}\,;
\,\,\,\,
 {B^Q_l}^*= {{{B^\Omega _l}^*} \over {1+{B^\Omega _l}^*}}\,. 
\label{PV1}             
\end{equation} 
Eq.\ (\ref{PV1}) shows that
the standard results of the microscopic FLT are recovered if one identifies
the Landau parameters with the FP values of $A^\Omega _l$ and $B^\Omega _l$:
\begin{equation}
F^s_l={A^\Omega _l}^*\,;\,\,\,\, 
F^a_l={B^\Omega _l}^*\,.
\label{PV2}
\end{equation}
Alternatively, (\ref{PV2}) can be written as $f_{\sigma _i}(\theta )={\Gamma
^\Omega _{\sigma _i}}^*(\theta )$ where $f_{\sigma _i}(\theta )$ is Landau's
quasi-particle interaction function.  
Since the singular contribution (\ref{ZS'2}) of the ZS$'$ graph to $\Gamma
_{\sigma _i}^\Omega $ was neglected when approximating (\ref{RGEQ2}) by
(\ref{RGEQ3}), 
${\Gamma _{\sigma _i}^Q(\theta )}^*$ obtained from (\ref{PV1}) is correct
only for $\vert \theta \vert \gtrsim T/E_F$. The determination of ${\Gamma
_{\sigma _i}^Q(\theta )}^*$ for $\vert \theta \vert \lesssim T/E_F$ would
require the consideration of the singular contribution of the ZS$'$ graph.
It should be noted that the diagrammatic derivation of FLT
\cite{Landau59,Abrikosov63} also neglects  the zero-angle singularity in 
the ZS$'$
channel and therefore does not respect the antisymmetry of the ``physical''
limit of the forward scattering vertex. The condition $\Gamma _t^Q(\theta
=0)=0$ is usually enforced,  giving the ``amplitude sum rule'' of
FLT. \cite{Baym91} For physical quantities (like
collective modes or response functions) which probe all values of the angle
$\theta $, it is nevertheless justified to neglect the singularity of the
ZS$'$ channel. 

The relation (\ref{PV1}) between ${\Gamma ^Q}^*$ and the Landau parameters has
been obtained at one-loop order. It appears therefore as an approximate
relation whose validity is restricted to the weak coupling limit. However,
it turns out that (\ref{PV1}) holds at all orders in a loop
expansion if we assume that the only singular contribution 
to the RG flow comes from the one-loop ZS graph (again we
neglect the singular contribution of the ZS$'$ graph). Note that the same kind
of assumption is at the basis of the diagrammatic 
derivation of FLT. \cite{Landau59,Abrikosov63} In this case, the RG flows
of $\Gamma ^Q$ and $\Gamma ^\Omega $ are determined by
\begin{eqnarray}
{{d\Gamma ^Q_{\sigma _i}} \over {dt}} &=& 
{{d\Gamma ^Q_{\sigma _i}} \over {dt}} \Biggl \vert _{{\rm 
ZS}} +
{{d\Gamma ^Q_{\sigma _i}} \over {dt}} \Biggl \vert 
_{{\rm ZS}',{\rm BCS},{\rm 2\, loops...}} 
\label{RGEQ4} \\ 
{{d\Gamma ^\Omega _{\sigma _i}} \over {dt}} &=& 
{{d\Gamma ^\Omega _{\sigma _i}} \over {dt}} \Biggl \vert 
_{\rm  {\rm ZS}',{\rm BCS,2\, loops...}} =
{{d\Gamma ^Q_{\sigma _i}} \over {dt}} \Biggl \vert 
_{\rm ZS',BCS,2\, loops...}  
\label{RGEQ5}
\end{eqnarray} 
The contribution of the one-loop ZS graph (first term of the rhs of
(\ref{RGEQ4})) has been separated from the non-singular contributions. Note
that we have included in this latter the one-loop BCS graph which had been
neglected up to now. Eqs.\ (\ref{RGEQ4},\ref{RGEQ5}) can be
combined to obtain 
\begin{equation}
{{d\Gamma ^Q_{\sigma _i}} \over {dt}}=
{{d\Gamma ^\Omega _{\sigma _i}} \over {dt}} + 
{{d\Gamma ^Q_{\sigma _i}} \over {dt}} \Biggl \vert _{{\rm 
ZS}} \,,
\label{RGEQ6}
\end{equation}
where  the second term of the rhs of (\ref{RGEQ6}) is given by (\ref{ZS1}).   
Since, according to our assumption, $\Gamma ^\Omega _{\sigma _i}$ is a
non-singular function of $\Lambda (t)$, Eq.\ (\ref{RGEQ6}) is similar to Eq.\
(\ref{RGEQ1}) and can be solved in the same way, yielding again the result 
(\ref{PV1}). Thus, higher order contributions change the FP
value ${\Gamma _{\sigma _i}^\Omega }^*$ obtained at one-loop order, but not
the relation (\ref{PV1}) between  ${\Gamma _{\sigma _i}^Q}^*$ and
${\Gamma _{\sigma _i}^\Omega }^*$. 

It should be pointed out that the Landau parameters are not determined by
the bare coupling function $U_{\sigma _i}({\bf K}_1,...,{\bf K}_4)$ of
the effective action (\ref{action}), but are related to the FP
value ${\Gamma ^\Omega _{\sigma _i}}^*$. Usually,
the FP values of physical quantities are related to some bare
effective values, so that their calculation by means of the RG (within the
framework of a loop expansion) is approximative and valid only in
the weak coupling regime. The FLT, which  assumes that the
only  singular contribution to the RG flow is due to the one-loop ZS graph,
does not rely on any kind of weak coupling condition. 

Eq.\ (\ref{RGEQ3}) (together with the analog Eq.\ for $B^Q_l$) 
is nothing else but the Bethe-Salpeter equation in the ZS
channel for the vertex $\Gamma ^Q_{\sigma _i}$, with 
${\Gamma ^\Omega _{\sigma _i}}^*$ the irreducible two-particle
vertex. This shows that the integration of the RG equations generates the
same Feynman diagrams as those considered by Landau.
\cite{Landau59,Abrikosov63} From this point of view, there is therefore a 
strict equivalence between the present RG approach and the diagrammatic
microscopic derivation of FLT. 

We would like to thank C. Bourbonnais, H. Schulz,
D. S\'en\'echal, A.-M. Tremblay Y. Vilk, 
and D. Zanchi for stimulating conversations and helpful comments. 
N.D. wishes to thank C. Bourbonnais and the condensed-matter group of the
university of Sherbrooke for hospitality. The Laboratoire de Physique des
Solides is Unit\'e Associ\'ee au CNRS. G.C. was supported by NSERC and by
FCAR. 

\begin{figure}
\epsfysize 4.5cm
\epsffile[10 260 500 570]{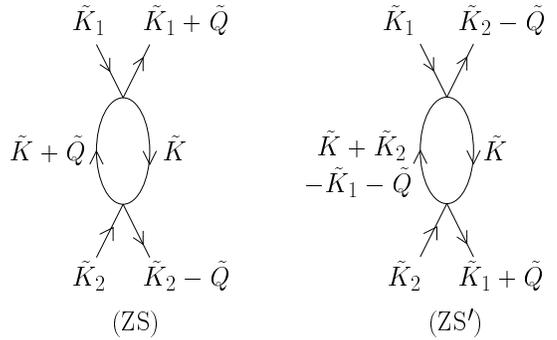}
\caption{One-loop diagrams for the renormalization of the vertex $\Gamma
_{\sigma _i}$ in the ZS and ZS$'$ channels (the spin indices are not
shown).  }
\end{figure}

\end{document}